# Modeling Corporate Epidemiology

Benjamin N. Waber[1], Ellen Brooks Pollock[2], Manuel Cebrian[1], Riley Crane[1], Leon Danon[3], and Alex (Sandy) Pentland[1]
[1]MIT Media Laboratory, [2]London School of Hygiene and Tropical Medicine,
[3]University of Warwick

Corporate responses to illness is currently an ad-hoc, subjective process that has little basis in data on how disease actually spreads at the workplace. Some businesses emphasize a "tough-it-out" approach, implicitly encouraging employees to remain at work while they are sick. Other companies send people home at the first symptom and force them to stay home long after any symptoms remain. Naturally there is a continuum of responses, but the appropriateness of these responses has not been gauged. This problem is not trivial, with the common cold costing the US economy close to $30 billion in lost productivity [1].

Particularly troubling is that many studies have shown that productivity is not an individual factor but a social one. Large scale studies on the economic impact of disease have only considered the productivity lost when an individual has to miss work or their reduced effectiveness if they remain at work [2]. In previous work with this dataset we found that a one standard deviation increase in network constraint in face-to-face networks was associated with a 10% increase in productivity [11], implying that if opportunities for face-to-face interaction were limited due to illness, productivity may suffer. Recently, in a causal study we showed that by modifying employee break structure we were able to increase network constraint and markedly increase productivity [10]. Therefore in any study on epidemic responses this social factor has to be taken into account.

The chief barrier to addressing this problem has been the lack of data on the interaction and mobility patterns of people in the workplace. Surveys are inaccurate and human observation is difficult and limited in scope [3, 4]. The emergence of low-cost sensing technology as a scientific tool has changed this equation, however, and researchers have begun using this data to examine questions that before were thought intractable.

We have created a wearable Sociometric Badge that senses interactions between individuals using an infra-red (IR) transceiver and proximity using a radio transmitter [9]. Face-to-face interactions are a major vector for disease transmission, particularly for common diseases such as influenza and the common cold [8]. Using the data from the Sociometric Badges we are able to simulate diseases spreading through face-to-face interactions with realistic epidemiological parameters. In terms of air exchange proximity is a poorer measure than interactions at detecting probability of infection, but proximity is an easier to obtain measure due to the widespread availability of Bluetooth proximity data. In this paper we also compare proximity-based infections to interaction-based infections to determine if proximity data is suitably fine-grained for modeling purposes.

In this paper we aim to construct a curve trading off productivity with epidemic potential. This is advantageous because it will allow companies to decide appropriate responses based on the organizational context of a disease outbreak. Compared with other studies, we are able to take into account impacts on productivity that arise from social factors, such as interaction diversity and density, which studies using an individual approach ignore. We also propose new organizational responses to diseases that take into account behavioral patterns that are associated with a more virulent disease spread. We believe that as sensing data becomes more commonplace in organizational and epidemiological studies [1, 5, 7] there will be a fast uptick in the number and sophistication of disease responses, and we hope that the approach we outline here will be used as a blueprint.

**Study Description**

We deployed our Sociometric badge platform for a period of one month (20 working days) at a one year old division of a Chicago-area data server configuration firm that consisted of 56 employees with 36 participating in the study.

Each employee was instructed to wear a Sociometric badge every day from the moment they arrived at work until they left their office. The salesman in the field used an automated program to request a computer system configuration for a potential customer. These configurations are automatically assigned a difficulty (basic, complex, or advanced, in ascending order of difficulty) based on the configuration characteristics. Employees in the department are then assigned a configuration task in a first come first served fashion. This configuration task may require them to use a computer aided design program in order to satisfy the customer's needs. Finally, the employee submits the completed configuration as well as price back to the salesman, and the employee is placed at the back of the queue for task assignment. The exact start and end time of the task is logged, and the number of follow-ups that are required after the configuration is completed is also recorded in the database. We used (negative) completion time as our measure of productivity, since shorter completion times are more desirable, and in this organization employees are rewarded based on their throughput.

**Transmission Model**

Using the data collected over the period of the study, we defined a stochastic, individual based model of disease transmission [6]. Each pair-wise interaction is assumed to carry a probability of disease transmission, should one individual of the pair be infected and therefore shedding disease. The probability of disease transmission scaled with the duration of the interaction with a rate of 0.007 per minute and a recovery rate of 1/3 per day. For simplicity, we assume that the incubation period was nil, such that as soon as individuals became infected, they become infectious, and remain infections at the same level until they recover.

We measure the epidemic potential in terms of the cumulative number of individuals infected (final epidemic size). Since the model is stochastic, the final size is averaged over many simulations to obtain meaningful results. Simulations are repeated starting with each individual in turn and at the beginning of each day during the first week to model different routes of introduction of disease. Each point is averaged over 185 repetitions.

The baseline final size assumes that all contacts occurred irrespective of infection status. To mimic potential organizational responses to epidemic threat, we eliminated interactions of a certain length of time in two ways: a) interactions with duration below a given threshold were eliminated and b) interactions with duration above a given threshold were eliminated.

Changes in network constraint after interactions are removed and used to predict changes in productivity using the results from [11] (i.e. a one standard deviation change in cohesion implies 10% higher productivity).

**Preliminary Results**

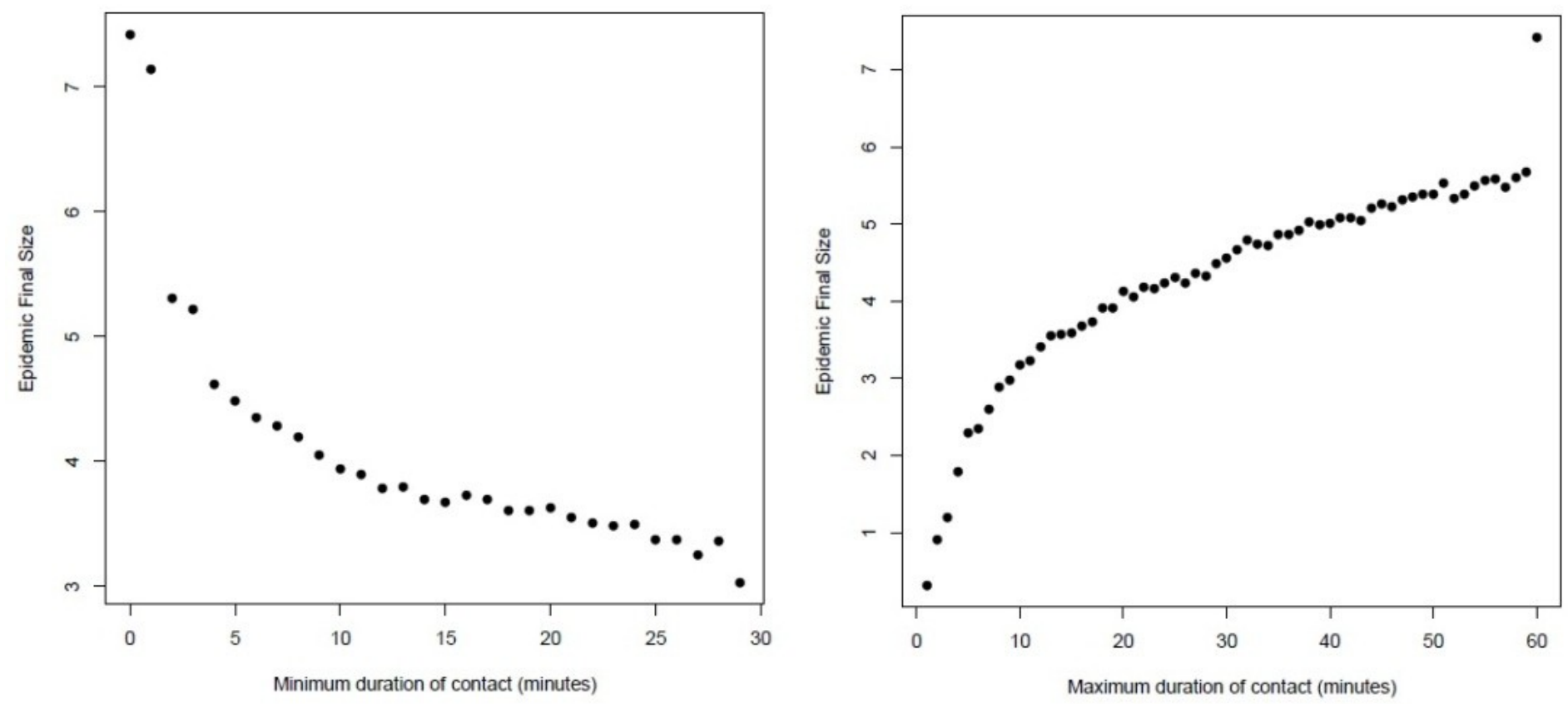

Figure 1: Epidemic Final Size

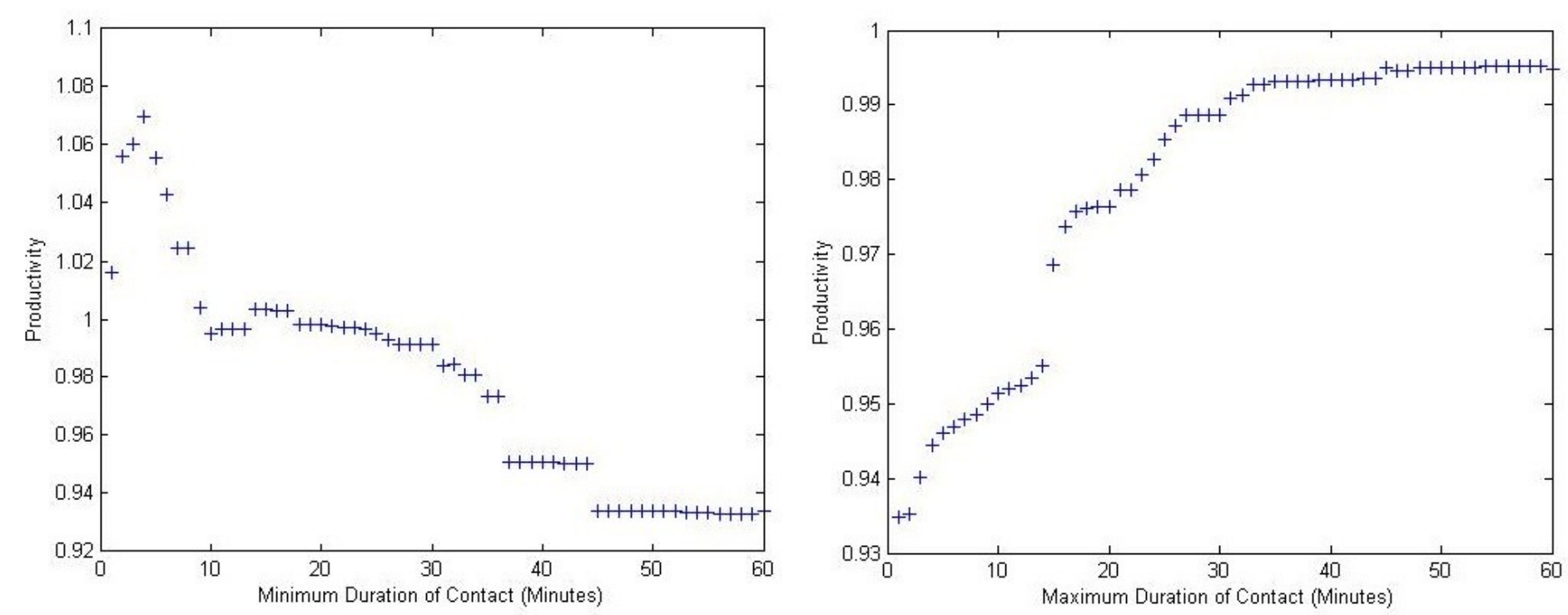

Figure 2: Productivity Changes. 1 represents the original department-level productivity.

In the baseline model with all interactions permitted, just over 7 individuals were infected during the course of the epidemic. We found that eliminating interactions of less than x minutes reduced the mean final size (figure 1, left panel), however the effect was minimal when only interactions of less than a minute were eliminated. Our results suggest that eliminating interactions of less than 5 minutes can cut the final epidemic size by nearly 50%. Our data shows that removing these interactions may actually improve performance by over 5% due to the increase in constraint, and productivity does not drop more than 1% until all interactions below 30 minutes are removed (figure 2, left panel). Therefore it may be advantageous for organizations to declare high-risk infection periods where steps are taken so that people spend more time at their desk and reduce casual interactions.

From these results we can also see that removing all lengthy interactions (over 30 minutes) has a protective effect, although the relative reduction is less dramatic (figure 1, right panel). This has a consistently negative effect on performance, however productivity does not drop by more than more than 1% until all interactions longer than 26 minutes are removed since these interactions have a negligible effect on constraint (figure 2, right panel). This suggests that while long meetings are infrequent, they are responsible for a large proportion of epidemic spreading in the workplace, and rather reduce casual interactions a similar effect could be obtained by delaying meetings to a low-risk period.

It remains to relate these potential organizational responses to productivity changes. We also compare taking a sick day versus coming to work and risk infecting others by incorporating productivity information. For the workshop we will also present that portion of the analysis and highlight additional results. From our findings it is clear that this study will have profound implications for how organizations respond to illnesses, and we hope that this will spur further research in this area in the future.